\documentclass[aps,pra,showpacs,twocolumn,superscriptaddress]{revtex4-1}
\usepackage{bm,color,amsmath,amssymb,mathrsfs,latexsym,graphicx,psfrag}

\usepackage{ifthen}
\newboolean{insert_titles}
\setboolean{insert_titles}{false} 
\newcommand{\opt}[1]
{
  \ifthenelse{\boolean{insert_titles}}{#1}{}
}

\newcommand{\ket}[1]{|#1\rangle}

\newcommand{\ev}[1]{\langle #1 \rangle}

\newcommand{\bk}[0]{\mathbf k}

\newcommand{\spl}[1]{\begin{align}\begin{split} #1 \end{split} \end{align}}

\begin{document}

\title{Creating exotic condensates via quantum-phase-revival dynamics in engineered lattice potentials}

\author{Michael Buchhold}
\affiliation{Institut f\"ur Theoretische Physik, Johann Wolfgang Goethe-Universit\"at, 60438 Frankfurt/Main, Germany}
\author{Ulf Bissbort}
\affiliation{Institut f\"ur Theoretische Physik, Johann Wolfgang Goethe-Universit\"at, 60438 Frankfurt/Main, Germany}
\author{Sebastian Will}
\affiliation{Fakult\"at f\"ur Physik, Ludwig-Maximilians-Universit\"at, 80799 M\"unchen, Germany}
\affiliation{Max-Planck-Institut f\"ur Quantenoptik, 85748 Garching, Germany}
\author{Walter Hofstetter}
\affiliation{Institut f\"ur Theoretische Physik, Johann Wolfgang Goethe-Universit\"at, 60438 Frankfurt/Main, Germany}

\begin{abstract}
In the field of ultracold atoms in optical lattices a plethora of phenomena governed by the hopping energy $J$ and the interaction energy $U$ have been studied in recent years. However, the trapping potential typically present in these systems sets another energy scale and the effects of the corresponding time scale on the quantum dynamics have rarely been considered.  Here we study the quantum collapse and revival of a lattice Bose-Einstein condensate (BEC) in an arbitrary spatial potential, focusing on the special case of harmonic confinement. Analyzing the time evolution of the single-particle density matrix, we show that the physics arising at the (temporally) recurrent quantum phase revivals is essentially captured by an effective single particle theory. This opens the possibility to prepare exotic non-equilibrium condensate states with a large degree of freedom by engineering the underlying spatial lattice potential.
\end{abstract}


\maketitle

\section{Introduction}
The progress in the field of ultracold atoms has opened a new path towards directly observing quantum dynamics in the laboratory. The interference of condensates, Bloch oscillations, solitons and dynamics of spinor condensates are examples of interesting dynamical effects, which, prior to their direct realization in cold atomic systems were hard to observe and only acted as theoretical textbook examples \cite{Bloch2008, Stoferle2004, Spielman2007, Greiner2002a, Greiner2002b}. The first step was the realization of non- or weakly interacting condensates, which allowed for the observation of single particle quantum dynamics \cite{Dalfovo1999, Denschlag2000, Matthews1999, Khaykovich2002}. However, the more interesting systems are those with strong inter-particle interactions, beyond the scope of a single particle Gross-Pitaevskii description. Strongly interacting systems became available using the capabilities of Feshbach resonances and optical lattices. An intriguing example for strongly correlated quantum dynamics is collapse and revival (CR) of a condensate when the lattice is suddenly ramped up and local interactions dominate \cite{Greiner2002b, Anderlini2006, Kollath2007, SebbyStrabley2007}. This has been successfully used to observe physics beyond the Bose-Hubbard model, quantifying the density-dependence of the intra- and interspecies interaction energy arising from the admixture of higher band contributions \cite{Will2010, Johnson2009, Will2010b, Mering2011}. In recent works \cite{Fischer2008,Wolf2010,Schachenmayer2011} the effect of a trapping potential on the $n(\bk=0)$ component during CR, as well as a renormalization of the revival time by a finite $J$ have been considered. 

\begin{figure}[t!]
\includegraphics[width=0.9\linewidth]{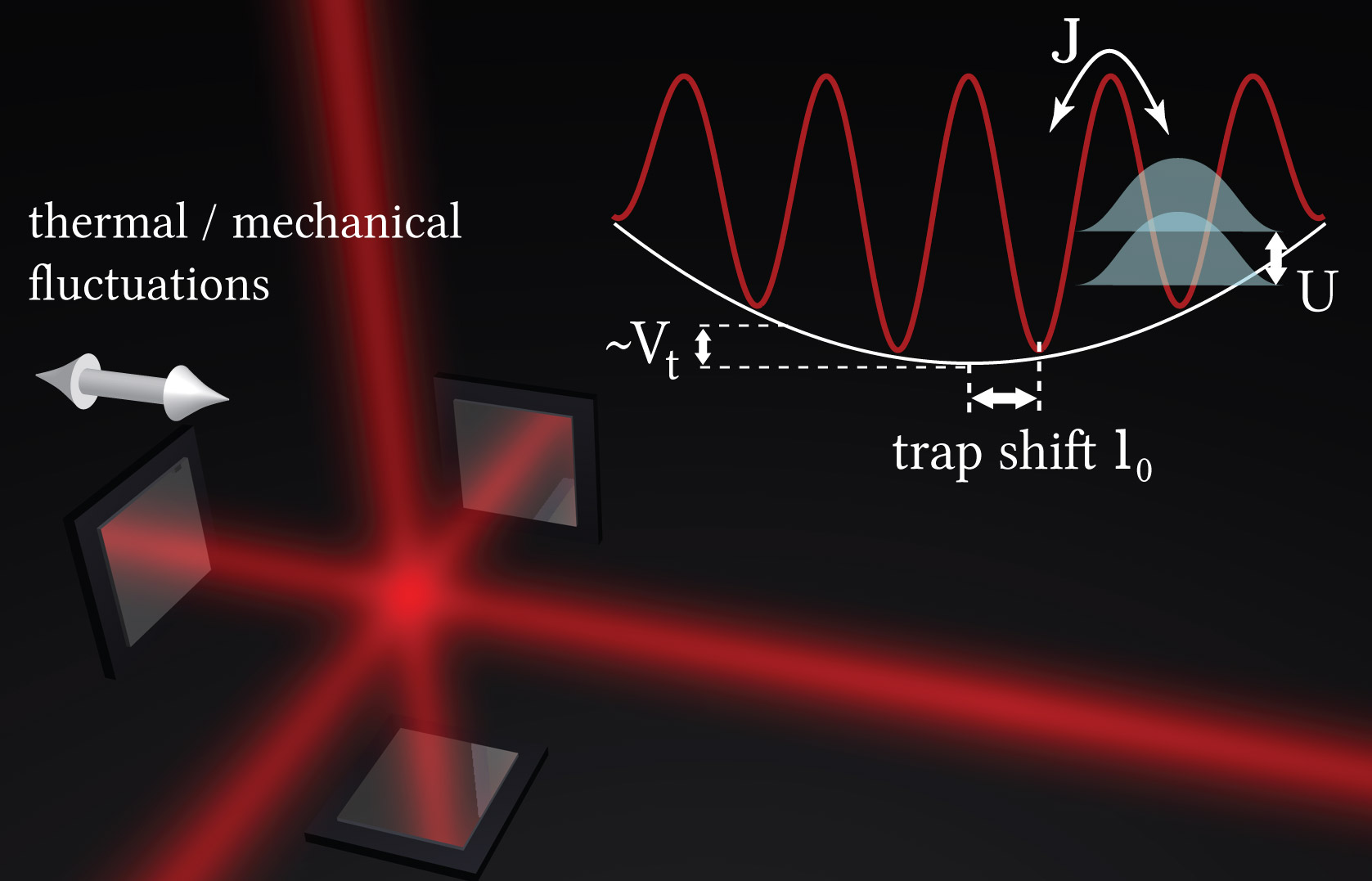}
    \caption
    {\label{fig:laser_setup}
(Color online). The trap shift ${\mathbf{l}}_0$ is defined as the distance between the minimum of the harmonic potential and the center of the next lattice site (in units of the lattice constant $a$). As the confinement is created by laser beams different from those creating the lattice in a given direction, thermal and mechanical fluctuations in the experimental setup can lead to a change in $\mathbf{l}_0$.
}
\end{figure}

Here we study the full momentum distribution during CR dynamics. We show that after a sudden lattice quench, the dynamics in an inhomogeneous spatial potential can be described by an effective single-particle theory at the discrete times of quantum phase revivals in the limit of negligible $J$. For the specific case of a harmonic potential, condensate states consisting of coherent superpositions of discrete, equidistantly spaced quasi-momentum states are obtained, forming an intriguing case of non-equilibrium state preparation. The position of the quasi-momentum components in the first Brillouin zone is highly sensitive to a shift of the trapping potential relative to the lattice (see Fig.~\ref{fig:laser_setup}) and the appearance of the peaked momentum pattern requires the harmonic trapping time scale to be a rational multiple of the revival time. In an experiment this allows to extract the global trapping potential with high precision. We derive an approximate analytic theory based solely on the dynamical evolution of the single-particle density matrix (SPDM). We verify the applicability under experimental conditions by furthermore performing a numerical analysis based on dynamic bosonic Gutzwiller theory, which includes condensate depletion in the initial state, finite size effects and finite tunneling.

We briefly review the basic CR physics, and proceed by extending the analysis to inhomogeneous systems, including other effects, such as finite tunneling and density-dependent interaction parameters. To observe CR, a bosonic system of atoms in an optical lattice is prepared in a Bose condensed state, before the lattice depth $s$ (expressed in units of the recoil energy $E_r$) is suddenly ramped up \cite{Will2010,Will2010b,Greiner2002b} (typically to $s\geq 25$), such that the interaction strength $U$ becomes the dominating energy scale. 

\section{Homogeneous Case}
In the case of a homogeneous system, where all lattice sites are equivalent and $s$ is sufficiently high that $J/U$ can be neglected \cite{small_J_over_U}, the subsequent time evolution is generated by the Hamiltonian $\mathcal H=\frac U 2\hat n( \hat n -1)$. The evolution of a local annihilation operator can be expressed in the Heisenberg representation as (choosing $\hbar=1$)
\begin{equation}
\label{eq:b_time_ev}	
	b(t)=e^{i \mathcal H t}b \,e^{-i \mathcal H t}=e^{-i U \hat n t} \, b	\, .
\end{equation}
We note that the bosonic Gutzwiller method is well justified here, as it assumes a product state of local on-site states. The local number statistics in the superfluid (SF) before the ramp-up, including number squeezing at initially finite $U/J$ is well described and becomes poissonian in the limit $U/J \to 0$. In this ideal limit, the ground state for a system consisting of $N$ particles on $L$ sites becomes a coherent state in the $\bk=0$ mode $\ket{\psi}=\ket{z}_1 \otimes \ldots \otimes \ket{z}_L$ within a grand-canonical, $U(1)$-symmetry breaking description with a density $n=N/L=|z|^2$. For this state the expectation value of the order parameter (\ref{eq:b_time_ev}) during CR can then be evaluated exactly \cite{Will2010, Imamoglu1997, Castin1997}
\begin{align}
\begin{split}
\label{Eq:homogeneous_psi_t_dep}
\psi(t) =\ev{b(t)}= z\exp\left(|z|^2\left(e^{-i{Ut}}-1\right)\right)\, .
\end{split}
\end{align}

The condensate fraction is defined as $f_c=\lambda_0/N$, where $\lambda_0$ is the largest eigenvalue of the SPDM $\rho_{ \mathbf{l},\mathbf{l}'}=\ev{  b_{\mathbf{l}}^\dag b_{\mathbf{l}'}^{\phantom{\dag}}}$ and $\mathbf{l}$ is the vectorial site index on the 3D lattice. In the limit of a large homogeneous system, Eq.~(\ref{Eq:homogeneous_psi_t_dep}) leads to
\begin{equation}
\label{Eq:homogeneous_cond_density_t_dep}
f_c(t)=e^{2|z|^2\left(\cos\left({Ut}\right)-1\right)},
\end{equation}
which is monotonically related to the visibility measured in experiments \cite{Gerbier2005, Will2010}. At times $t_m= m\ t_{\mathrm{rev}}$ with the revival time $t_{\mathrm{rev}}=2\pi/U$ and $m\in \mathbb N$, the coefficients $c_n(t)$ in the Fock representation $\ket{\psi(t)}_{\mathbf{l}}=\sum_{n=0}^\infty c_n(t) \ket{n}_{\mathbf{l}}$ periodically coincide after having performed $mn(n-1)/2$ rotations in the complex plane, which leads to a revival of the initial condensate.
In the case of finite interactions $U/J>0$ prior to the ramp-up, the phenomenological ansatz 
\begin{equation}
\label{cond_frac_homogeneous_interacting}
f_c(t)=\alpha \,e^{2\beta\left(\cos\left({Ut}\right)-1\right)}-\gamma
\end{equation}
leads to a remarkably good agreement with the exact numerical results for arbitrarily long times.
The fitting parameters $\alpha,\beta,\gamma \in \mathbb R$ incorporate number squeezing and depletion in the initial condensate.
\begin{figure}[b!]
\begin{center}
\includegraphics[width=\linewidth]{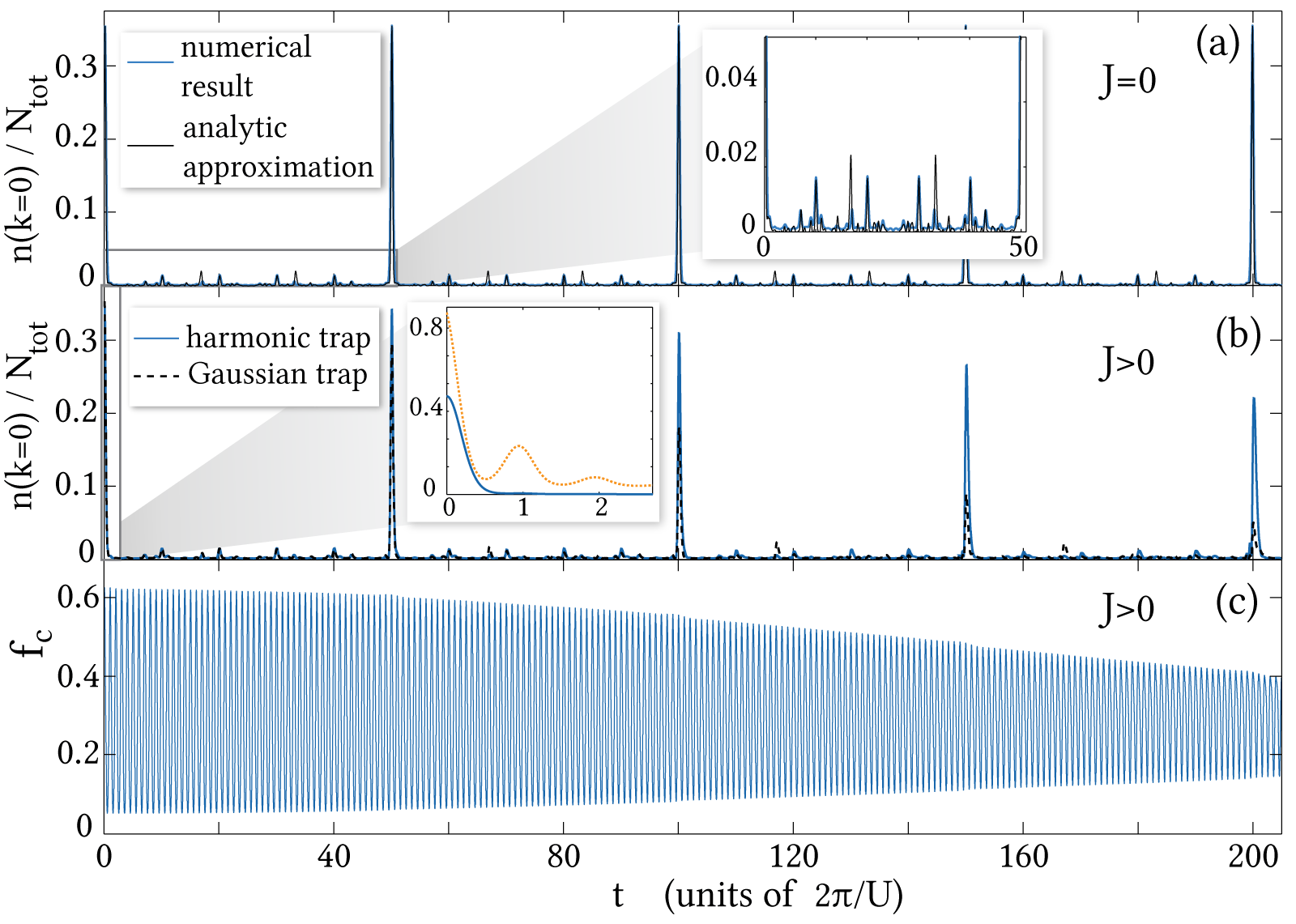}
    \caption
    {\label{fig:revival_fct_time}   
(Color online). Time-dependent occupation of the $\bk=0$ mode (a, b) and condensate fraction (c) for a $^{87}$Rb condensate prepared in a $738\mbox{nm}$ lattice with a central density of $n=2.6$. The lattice is ramped up from $s=8$ and a $157.4$Hz trap to $s=26$ and a $346.2$Hz trapping frequency with $\mathbf{l}_{0,i}=0.5$, such that $t_{\mathrm{trap}}=100t_{\mathrm{rev}}=28.29$ms. In subplot (a), $J$ is artificially set to zero after the ramp to compare the numerical results to Eq. \ref{n_of_k_approx}.  The inset in (b) shows the initially strong decline in the occupation of the $\bk=0$ mode (blue line) and the central peak $N_p(t)$ (dotted orange line), see \cite{N_central_peak}. Additionally, the dashed line in panel (b) shows $n(\bk=0,t)$ for a Gaussian trap, with parameters corresponding to the harmonic potential, and a cloud extension ratio $\frac{R_{cloud}}{w}=0.1$.}
\end{center}
\end{figure}

\section{Inhomogeneous Case}
For an inhomogeneous system without translational symmetry, the single site description does not suffice. The time evolution in the atomic limit (neglecting $J/U$) is generated by the lattice Hamiltonian $\mathcal H=\frac U 2 \sum_{\mathbf{l}} \hat  n_{\mathbf{l}}( \hat n_{\mathbf{l}} -1) - \sum_{\mathbf{l}} \mu_{\mathbf{l}} \hat n_{\mathbf{l}}$ with the effective chemical potential $\mu_{\mathbf{l}}$ at lattice site ${\mathbf{l}}$ accounting for the inhomogeneity. Analogously, the time evolution of the local annihilation operators in the Heisenberg picture is
\begin{equation}
\label{eq:b_l_time_ev}
	b_{\mathbf{l}}(t)=e^{i \mathcal H t}b_{\mathbf{l}} \,e^{-i \mathcal H t}=e^{-i(U  \hat n_{\mathbf{l}}   -\mu_{\mathbf{l}})t} b_{\mathbf{l}}.
\end{equation}
In case of an initially ideal (but not necessarily homogeneous) condensate, the time- and position-dependent local order parameter can be evaluated exactly
\begin{equation}
\label{eq:psi_l_time_ev}
	\psi_{\mathbf{l}}(t)=\ev{b_{\mathbf{l}}(t)}=z_{\mathbf{l}} \, e^{i\mu_{\mathbf{l}} t} e^{|z_{\mathbf{l}}|^2(e^{-iUt}-1)}.
\end{equation}
We now turn to the condensate fraction by the convenient decomposition of the SPDM using 
Eq.~\ref{eq:b_l_time_ev}
\begin{align}
\begin{split}
	\rho_{{\mathbf{l}},{\mathbf{l}}'}(t)&=e^{-i \mu_{\mathbf{l}} t} \ev{b_{\mathbf{l}}^\dag \, e^{iU( b_{\mathbf{l}}^\dag  b_{{\mathbf{l}}}^{\phantom{\dag}} -b_{{\mathbf{l}}'}^\dag  b_{{\mathbf{l}}'}^{\phantom{\dag}}  )t} b_{{\mathbf{l}}'}^{\phantom{\dag}} } e^{i \mu_{{\mathbf{l}}'} t}.
\end{split}
\end{align} 
\begin{figure*}[t!]
\includegraphics[width=\linewidth]{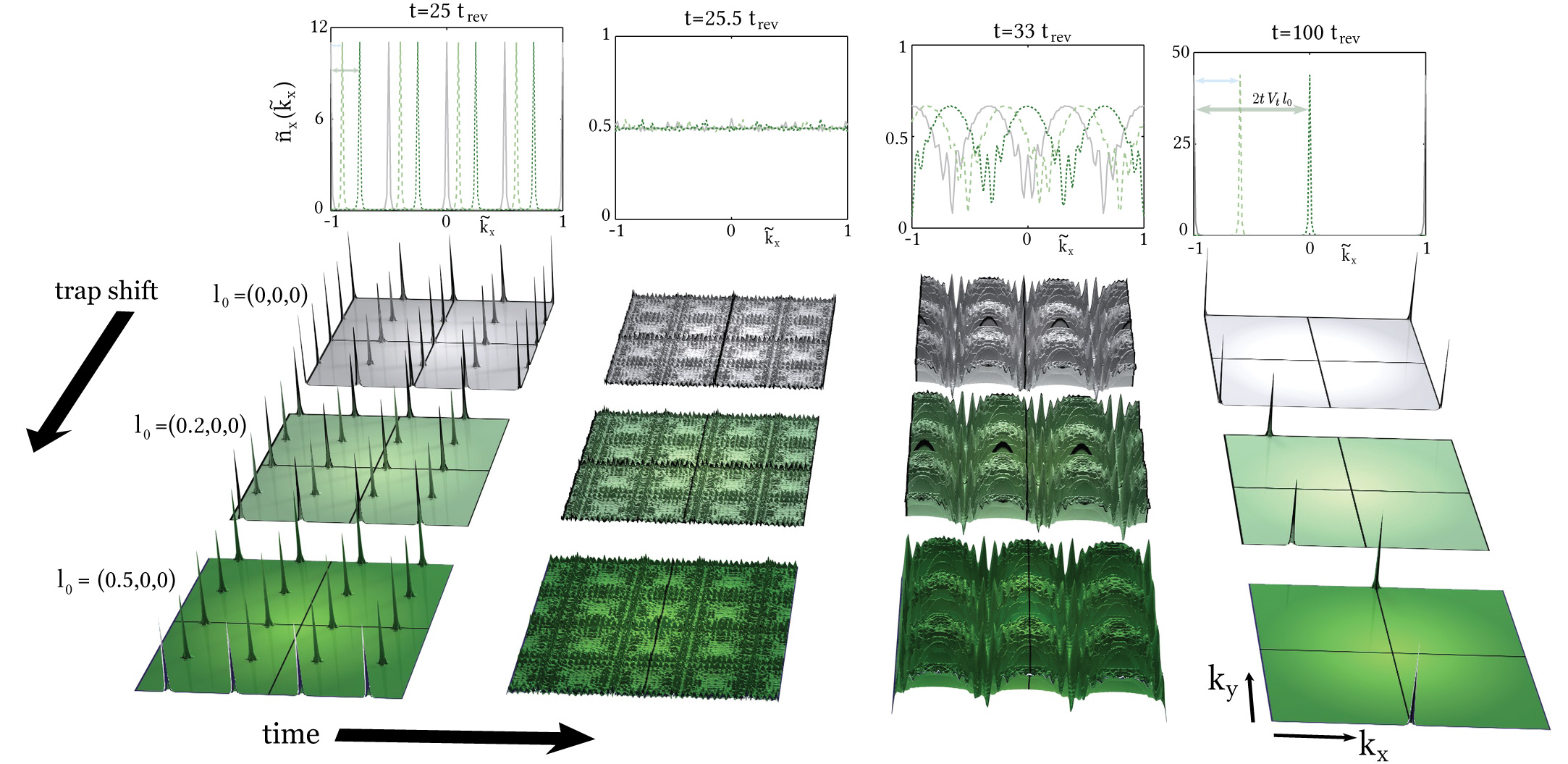}
    \caption
    {\label{momentum_dists}
    (Color online). Quasi-momentum profile in the first Brillouin zone for different times and lattice shifts ${\mathbf{l}}_0$ during CR in a harmonic trap, individually scaled on the vertical axis for visual clarity. Parameters as in Fig.~\ref{fig:revival_fct_time}, but initially in a $27.5$Hz trap with a central density of $n=1.67$. Subsequently, this is ramped up to $s=40$ and a $197.3$Hz trapping frequency, such that $t_{\mathrm{trap}}=200t_{\mathrm{rev}}=43.88$ms. Each of the plots in the upper row depicts the projected quasi-momentum density $\tilde n_x(k_x) \propto \int_{-\pi/a}^{\pi/a} dk_y \, dk_z \, n(\bk)$ for the three 3D plots (for a given time, but different trap shifts) in the respective column. These projected distributions are all normalized such that $\int_{-1}^{1} d \tilde k_x \, \tilde n(\tilde k_x)=1$, where $\tilde k_x= k_x a / \pi$. A trap shift of $l_0$ in the $x$-direction does not change the overall structure of $n(\bk)$, but leads to a translation of the quasi-momentum profile by exactly $2 t V_t l_0$ in the $k_x$-direction. This shift is indicated by the arrows in the projected quasi-momentum profiles in the upper row.
}
\end{figure*}
Remarkably, the time-dependent condensate fraction $f_c(t)$ in an inhomogeneous system is identical to that in the homogeneous system (i.e. $\mu_{\mathbf{l}}=\mu$) for a sufficiently deep lattice (for $J/U\approx 0$ to hold) and identical $U$. Here, the SPDM $\rho(t)=U(t)\, \tilde \rho(t) \, U^\dag(t)$ is connected to the SPDM of the homogeneous system during CR at the same time $\tilde \rho_{{\mathbf{l}},{\mathbf{l}}'}(t)=\ev{b_{\mathbf{l}}^\dag \, e^{iU( b_{\mathbf{l}}^\dag  b_{{\mathbf{l}}}^{\phantom{\dag}} -b_{{\mathbf{l}}'}^\dag  b_{{\mathbf{l}}'}^{\phantom{\dag}}  )t} b_{{\mathbf{l}}'}^{\phantom{\dag}} }$ through a time dependent unitary transformation by a diagonal matrix with elements $U_{{\mathbf{l}},{\mathbf{l}}'}(t)=e^{-i \mu_{\mathbf{l}} t} \delta_{{\mathbf{l}},{\mathbf{l}}'}$, thus leaving the eigenvalues and hence the condensate fraction invariant (we emphasize that this does not rely on any approximation). This equivalence means that the revival of the condensate is actually not damped by spatial inhomogeneities. At revival times the condensate state differs from the $\bk=0$ Bloch state, which appears as damping in certain observables such as the visibility \cite{Greiner2002b, Anderlini2006, SebbyStrabley2007}, but is fundamentally different from a real damping of the condensate. This relation also holds in the case of depleted condensates, where finite interactions are present prior to the ramp up, as also reflected by the numerical results for a harmonic trap shown in Fig.~\ref{fig:revival_fct_time}: whereas $f_c(t)$ (subplot (c)) is only damped by a finite $J$, $n(\bk=0)$ is almost completely suppressed after a single revival time by the very strong harmonic trap (not considering much longer times, where it may reappear as discussed later).

To calculate the condensate fraction for the specific structure of the SPDM within a Gutzwiller state, a highly efficient method was found to be the iteration of the self-consistency condition  for the largest eigenvalue $\lambda_0$ in the $i$-th iteration step

\begin{equation}
	\lambda_0^{(i+1)}=\lambda_0^{(i)}\sum_{\mathbf{l}} \frac{|\psi_{\mathbf{l}}|^2}{\lambda_0^{(i)}-n_{\mathbf{l}}+|\psi_{\mathbf{l}}|^2},
\end{equation}
obtained from a rearrangement of the eigenvalue equation (see appendix A for detailed discussion). It can be shown that on the interval $[\max_j(n_j-|\psi_j|^2),N_{\mbox{\tiny tot}}]$, the largest eigenvalue is the only attractive fixed point and rapid convergence is achieved. In the case of strong depletion, an adapted Lanczos algorithm is used.

Reverting to the Schr\"odinger picture of the SPDM, the decomposed form shows that if a condensate state (i.e.~a normalized eigenvector $\phi_{\mathbf{l}}^{(0)}(t)$ of $\rho(t)$ corresponding to the macroscopic eigenvalue) is present at time $t$, its time evolution is exclusively determined by the operator $U^\dag(t)$, i.e. $\phi_{\mathbf{l}}^{(0)}(t)=\sum_{{\mathbf{l}}'} U^\dag_{{\mathbf{l}},{\mathbf{l}}'}(t) \phi_{{\mathbf{l}}'}^{(0)}(0)$. 
This time evolution is formally also valid during collapse, where the condensate fraction is suppressed. Its observation, however, is only possible at revival times $t_m$, which can be tuned independently, thus effectively allowing the experimental observation of single particle dynamics in a lattice at arbitrary times. Note that although the effective single particle dynamics may resemble Gross-Pitaevskii theory, its derivation and validity is fundamentally different and relies on the opposite limit $U\gg J$.

Since the true condensate fraction is experimentally not directly accessible in inhomogeneous systems \cite{Gerbier2008, Moller2010}, we consider the dynamical evolution of the full quasi-momentum distribution $n(\bk)$. Under restriction to the lowest lattice band, the physical momentum distribution consists of replicas of the first Brillouin zone, scaled by the Wannier function's respective Fourier coefficient. Using the spectral decomposition of the SPDM $\rho_{{\mathbf{l}},{\mathbf{l}}'} = \sum_{i=0}^{L-1} \lambda_i \, {\phi_{{\mathbf{l}}}^{(i)}} {\phi_{{\mathbf{l}}'}^{(i)} }^*$ in conjunction with the defining property of a BEC, i.e. only a single macroscopic eigenvalue $\lambda_0$, $n(\bk)$ is found to be well approximated by
\begin{align}
\begin{split}
n(\mathbf{k},t)\approx& \frac N L f_c(t) \left|\sum_{\mathbf{l}} e^{-i{\mu_{\mathbf{l}}t}-ia{\mathbf{k}}{\mathbf{l}}}\phi_{\mathbf{l}}^{(0)}(0) \right|^2\hspace{-1.8mm}+C(t).
\label{n_of_k_approx}
\end{split}
\end{align}
To a good approximation, the non-condensed atoms lead to a constant incoherent background in $n(\bk)$ and $f_c(t)$ can be taken from Eq.~\ref{cond_frac_homogeneous_interacting}, exploiting its independence of spatial inhomogeneities during CR. The background term $C(t)=N(1-f_c(t))/L$ ensures particle number conservation, $\sum_\bk n(\bk)=N$.
We find that all qualitative features arising in a full time-dependent Gutzwiller simulation are captured surprisingly well by Eq.~\ref{n_of_k_approx} at all experimentally relevant time scales.

\section{Special Case: Harmonically Trapped System}
We shall now discuss the characteristics of this time evolution and possible applications. For the case of an underlying harmonic trapping potential $\mu_{\mathbf{l}}=-V_t({\mathbf{l}}-{\mathbf{l}}_0)^2+\mu_0$, results from a numerical dynamic Gutzwiller calculation for a trapped condensate (diameter\! $~\approx 120a$ in the $x$-$y$-plane) are shown in Fig.~\ref{fig:revival_fct_time} for $n(\bk=0,t)$ and the full $n(\bk,t)$ in Fig.~\ref{momentum_dists} for certain times. The essential dynamics of $n(\bk,t)$ at revivals is well captured within a single particle picture for the condensate state $\ket{\psi(t)}=\sum_{\mathbf{l}} \phi_{\mathbf{l}}^{(0)}(t) \ket {\mathbf{l}}$ (although the condensate vanishes between revivals), where $\ket{\mathbf{l}}$ is the Wannier state at site ${\mathbf{l}}$. Its time evolution is generated by $U(t)$, since the largest eigenvector of $\tilde \rho(t)$ is always identical at revival times.

The condensate's contribution to $n(\bk,t)$ (relevant during revivals) can then be expressed as 
\begin{equation}
	n(\bk,t)=f_c(t)\frac N L \sum_{{\mathbf{l}},{\mathbf{l}}'} A_{{\mathbf{l}},{\mathbf{l}}'} \, e^{-i(a {\mathbf{k}}-2tV_t {\mathbf{l}}_0)({\mathbf{l}}-{\mathbf{l}}')} \, e^{-it V_t({\mathbf{l}}^2-{\mathbf{l}}'^2)}
	\label{SP_momentum_profile}
\end{equation}
with the amplitudes $A_{{\mathbf{l}},{\mathbf{l}}'}=\phi_{\mathbf{l}}^{(0)}(0) {\phi_{{\mathbf{l}}'}^{(0)}}^*(0)$.
The first exponential term shows that a translation of the harmonic trapping potential by ${\mathbf{l}}_0$ leads to a time-dependent overall translation by $2tV_t {\mathbf{l}}_0$ of the quasi-momentum profile, which is periodic in the first Brillouin zone. The second term implies a temporal periodicity with a period of the trap time $t_{\mathrm{trap}}=2\pi/V_t$, up to a translation $4\pi n {\mathbf{l}}_0$.  At most times the last term interferes destructively, but at times $t_n^{(m)}=(n+1/m)t_{\mathrm{trap}}$ multi-peak structures appear in $n(\bk)$ and are observable if, additionally, $t_n^{(m)}$ is a multiple of $t_{\mathrm{rev}}$. In the limit of an initially homogeneous condensate, the discrete Fourier transform in Eq.~(\ref{SP_momentum_profile}) can be evaluated explicitly at the times $t_n^{(m)}$ (see Appendix B) and one obtains
\begin{figure}[t!]
\includegraphics[width=0.9\linewidth]{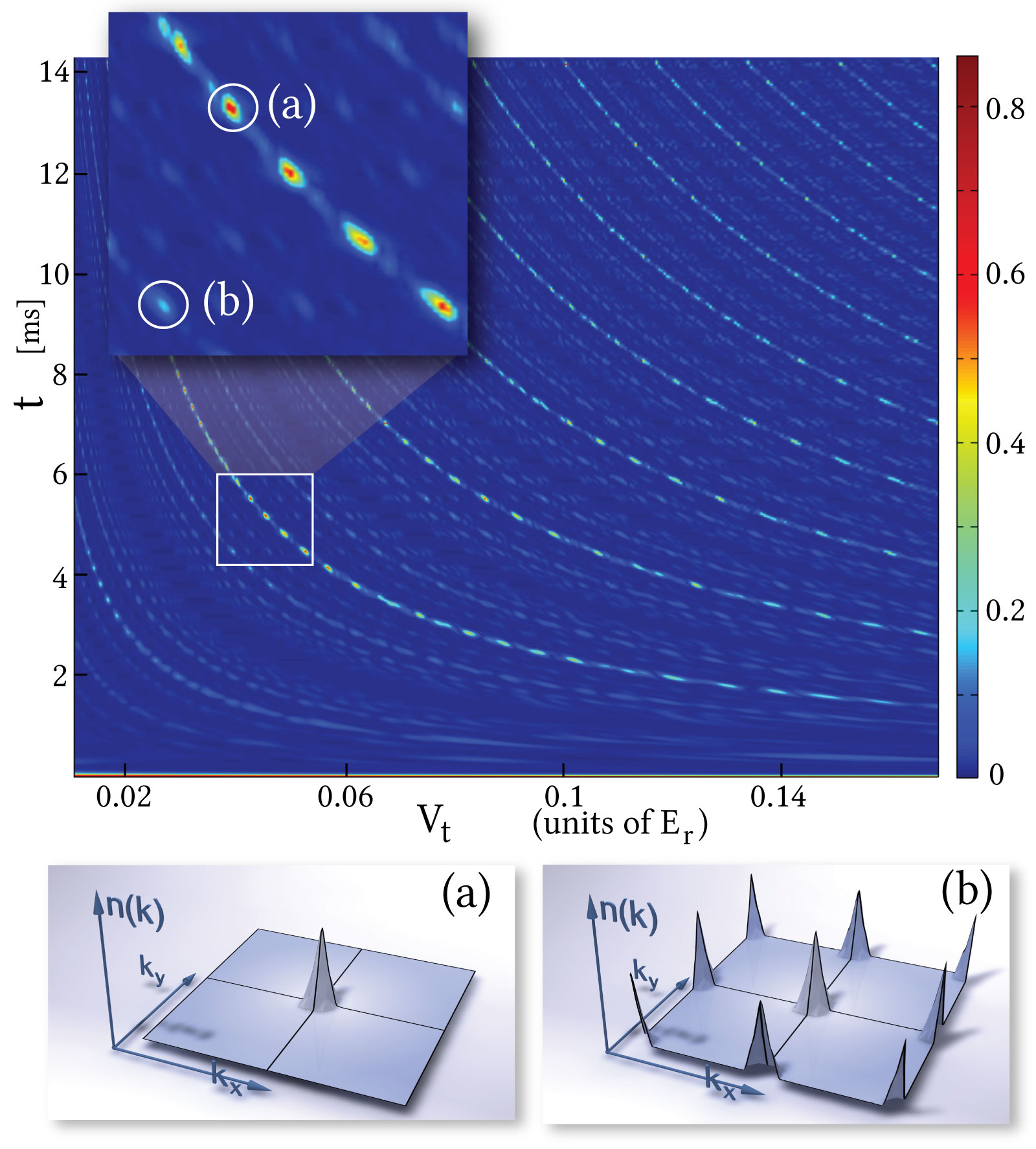}
    \caption
    {\label{fig:det_harmonic_trap_freq}
    (Color online). Particle number in the central quasi-momentum peak (see \cite{N_central_peak}) as a function of time and harmonic trapping strength $V_t$ for a trap shift ${\mathbf{l}}_{0,i}=0.5$. The position of the maxima allows for a high precision determination of $V_t$ in the presence of the optical lattice. The calculation was performed for $^{87}$Rb in a $738\mbox{nm}$ optical lattice at $s=32$, including finite $J$ effects and realistic density-dependent interaction parameters $U_n$ \cite{Will2010}. Insets: $n(\bk)$ at points (a),(b).
}
\end{figure}
\begin{equation}
\label{nk_of_t_res}
n(\bk,t)=N \, f_c(t)  \prod_{i=x,y,z}   \frac{F(q_i,m)}{m}
\end{equation}
with $F(q_i,m) = \delta_{q_i\bmod 1 ,\, 0      } \;\left( 1+\delta_{0,m \bmod 2} (-1)^{\frac m 2 +q_i  } \right)$ and  ${\bf q}=\frac{a m}{2 \pi}{\bf k} - 2 {\bf l}_0$. This constitutes the dominant contribution in the outermost left ($t_0^{(8)}$) and right columns ($t_0^{(2)}$) of Fig.~\ref{momentum_dists} during revival, whereas the peaks are strongly washed out in the second last column, where the revival time does not fully coincide with $(t_0^{(6)})$. During collapse, shown in the second column of Fig.~\ref{momentum_dists}, the distribution is essentially flat with little substructure.
We stress that at these specific times the condensed atoms collectively occupy one single particle state, consisting of a coherent superposition of quasi-momentum states $(K(m,\nu_x),K(m,\nu_y))$ with equal weight, i.e. for the 2D projected case in a mode
\begin{equation}
		a_{\mbox{\tiny BEC}}^\dag=\frac{1}{m}\sum_{\nu_x,\nu_y=1}^{m} a_{k_x=K(m,\nu_x),\,k_y=K(m,\nu_y)}^\dag
\end{equation}
with $\frac a \pi K(m,\nu_i)=\left[\frac 2 m (2(\nu_i+l_{0,i})+1 +\frac m 2) \right] \bmod 1$. This is not to be mistaken as a signature of a fragmented condensate.
As the appearance of these patterns is governed not only by the inverse trap time scale, but also by the CR time scale, the temporal discretization is implicitly given by the revival time $t_{\mathrm{rev}}$. A plot of $n(\bk=0,t)$ as a function of the trapping frequency is shown in Fig.~\ref{fig:det_harmonic_trap_freq}: the $1/V_t$ scaling of the time at which a pattern with a significant $n(\bk=0,t)$ component appears is in full accordance with Eq.~\ref{SP_momentum_profile} and the fainter hyperbolas correspond to momentum profiles with multiple spikes, where, correspondingly, the $k=0$ component is weaker. The non-persistent, dotted intensity of the hyperbolas directly reflects the interplay of the two time scales: only when a specific trap pattern coincides with a revival of the matter wave field, a significant $n(\bk=0)$ is observed. 
 
A further quantity that can be directly extracted from the distributions in Fig.~\ref{momentum_dists} at the revival times, is the position of the trapping potential in a scale smaller than the lattice spacing $a$. As can be seen from the term $-2tV_t {\mathbf{l}}_0$ in Eq.~\ref{SP_momentum_profile}, a microscopic shift ${\mathbf{l}}_0$ in the harmonic trapping potential leads to a time-dependent translation of the quasi-momentum distribution $n(\bk)$, as visible in the rows with different ${\mathbf{l}}_{0,x}=0,0.2,0.5$ in Fig.~\ref{momentum_dists}. This translation can be experimentally observed on a macroscopic scale after time-of-flight expansion. Thus by comparing the initial $n(\bk,t=0)$ with $n(\bk,t_m)$ at revival times, ${\mathbf{l}}_0$ can directly be determined with high precision. 

In turn, this may find application in characterizing the thermal or mechanical stability of the optical setup: as the confining harmonic potential is created by laser beams different from the ones generating the optical lattice in each dimension (see Fig.~\ref{fig:laser_setup}), thermal drifts \cite{thermal_fluct} can lead to a shift of $\mathbf{l}_0$, which can be observed in $n(\bk,t)$.

\begin{figure}[t!]
\includegraphics[width=\linewidth]{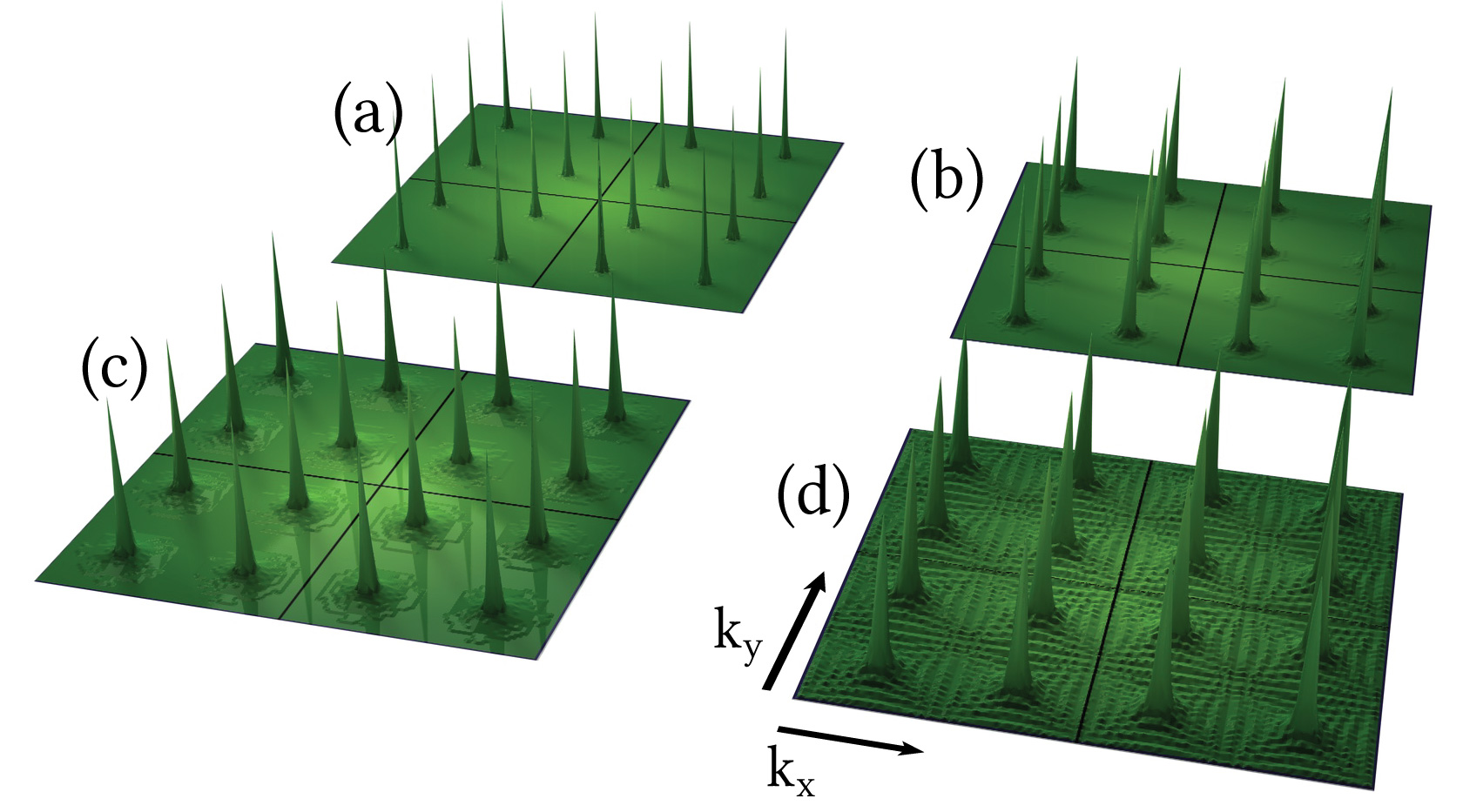}
    \caption
    {\label{fig:gaussian_comparison}
    (Color online). Comparison of the quasi-momentum profiles in the first Brillouin zone for quadratic (a) and Gaussian trapping potentials (b,c,d). The relevant parameter quantifying the harmonicity of the Gaussian trap is the ratio between the cloud radius and the width of the Gaussian potential (extension ratio). For subplots (b,c,d) this value was set to ($0.1$, $0.15$, $0.2$) respectively, whereas the trap strength and other parameters were set to agree with the strength of the harmonic potential in Fig.~3 in the linear approximation ($t=25t_{\mbox{\tiny rev}}$ and lattice shift ${\mathbf{l}}_0=(0.5,0.5,0.5)$). Compared to the harmonic trap, the peaks are successively broadened with an increasing filling factor for a Gaussian potential, but remain clearly visible, even for a large extension factor of $0.2$. Due to the scaling of the $\delta$-response of the peaks in $n(\bk)$ for the harmonic case, the theoretically predicted height of such a peak on a single point of the discrete $k$-lattice is strongly decreased (cf. black dotted line in Fig.~2b). In the experimental detection, this should be less visible, as the observed quasi-momentum distribution is in any case convolved with the point spread function of the optical imaging system.}
\end{figure}

\section{Validity of our predictions}
A crucial question is whether our predictions are robust enough to be observable for the harmonic trapping potential in typical experimental setups. As the most important effects we consider the sensitivity towards precise timing as well as deviations from the idealized harmonic trap shape due to the Gaussian intensity profile of the laser beam. For a given trap strength $V_t$, the required temporal precision (i.e. the time span between lattice ramp and the release of the cloud) is of the order of the revival time, as can be seen in Fig. 2 and Fig. 4. This time scale is precisely controlled in current experiments \cite{Will2010,Will2010b}. The deviations originating from the Gaussian trapping potential require a somewhat more quantitative analysis. The relevant parameter, which quantifies the quality of the harmonic approximation is the ratio of the cloud radius and the beam radius. In Fig. 2(b) we show the time-dependent occupation $n(\bk=0,t)$ for a true Gaussian trapping potential for the cloud extension ratio $\frac{R_{cloud}}{w}=0.1$ in comparison to the idealized harmonic case. We find that in this experimentally realistic regime the peaks are somewhat smaller, shorter and broader. However, on the time scale of $t_{\mbox{\tiny trap}}$ this effect is weak and should not wash out the structure of $n(\bk)$ beyond experimental resolution. This can also be seen in Fig. 5, where a comparison of the quasi-momentum profile after $t=25t_{\mbox{\tiny rev}}$ is given for a harmonic and Gaussian trap with different cloud extension ratios. In a very recent experiment \cite{Mark2011}, exactly these quasi-momentum patterns have been observed in the non-interacting one-dimensional system. In our picture, those data correspond to trap dynamics with an infinite revival time. Since the trapping potential in \cite{Mark2011} is not perfectly harmonic, the general feasibility of our predictions has already been experimentally confirmed.

The creation and observation of non-equilibrium condensate states in a non-interacting, one-dimensional system as reported in \cite{Mark2011} is contained in our theory as a limiting case. The quasi-momentum distribution of a corresponding one-dimensional system is nothing but the projection of the two-dimensional distribution on the $k_x$-axis. The emergent patterns in momentum space can be viewed as the temporal matter-wave analogon of the Talbot effect known from classical optics \cite{Talbot1, Talbot2}. From this point of view, one may also see our theory as a generalization of the temporal matter-wave Talbot effect to higher dimensional interacting systems (i.e. two or three dimensions).

\section{Conclusion}
In conclusion, we have analyzed the full momentum distribution during collapse and revival of an atomic condensate after a lattice quench, showing that the state at revival times can effectively be described by a single particle theory. We have shown that the true condensate fraction is not damped by the presence of an arbitrary spatial potential, but that the condensate state takes on a different form, such that a damping seems to appear when only looking at $n(\bk=0,t)$. For the specific case of a harmonic trapping potential, condensate states featuring a periodic peak structure in momentum space are created, allowing for high precision determination of the total trapping frequency of the underlying optical and magnetic trap, as well as the position of the trapping potential on a sub-lattice scale. By engineering the spatial potential after the ramp-up, our concept is suitable for the preparation of exotic non-equilibrium condensate states, as well as transforming different condensate states into each other. These prospects seem realistic in the light of recent experimental progress on the engineering of arbitrary spatial potentials \cite{Bakr2010, Sherson2010, Zimmermann2010}.

\begin{acknowledgments}
We thank I. Bloch and D. Semmler for fruitful discussions. This work was supported by the German Science Foundation (DFG) via Forschergruppe FOR 801.
\end{acknowledgments}

\begin{appendix}
\section{Efficient Calculation of the Condensate Fraction for a Bosonic Gutzwiller State}
To obtain the condensate fraction $f_c$ in an inhomogeneous system, it is necessary to determine the largest eigenvalue $\lambda_0$ of the SPDM $\rho_{l,l'}=\langle b^{\dagger}_l b^{\phantom{\dagger}}_{l'}\rangle$, which is a hermitian $L \times L$-matrix, $L$ being the number of lattice sites. The condensate fraction is then defined as $f_c=\frac{\lambda_0}{N_{\mbox{\tiny tot}}}$, where $N_{\mbox{\tiny tot}}$ is the total particle number. Since we determine $f_c(t)$ at many different times, an efficient algorithm to determine the largest eigenvalue is desirable. Here we present a method to determine the largest eigenvalue of the SPDM of a variational state of the bosonic Gutzwiller form, where the computational effort scales proportionally with $L$. We firstly derive relations holding for all eigenvalues of a SPDM of this form and subsequently use these to formulate a fix point relation for the eigenvalues. The largest eigenvalue can always be found by numerically iterating this fix point relation, as the largest eigenvalue on a large interval around the fix point.

We begin by considering the explicit form of the SPDM for a bosonic Gutzwiller state:
\begin{equation}
		\rho_{l,l'}=\langle b^{\dagger}_lb^{\phantom{\dagger}}_{l'}\rangle=\psi^*_l\, \psi_{l'}+\delta_{l,l'}\left(n_l-|\psi_l|^2\right)
\end{equation}
the off-diagonal part factorizes into products $\psi^*_l\, \psi_{l'}$ of local order parameters $\psi_{l}=\langle b^{\phantom{\dagger}}_l\rangle$, whereas the diagonal part consists of the local densities $n_l=\langle b^{\dagger}_lb^{\phantom{\dagger}}_{l}\rangle$.
Now let $\mathbf{v}$ be an eigenvector of $\rho$ with corresponding eigenvalue $\lambda$ (i.e. $\rho\mathbf{v}=\lambda\mathbf{v}$). Focusing on a single element $j$ of an eigenvector, we obtain
\begin{equation}
		\lambda\, v_j=\sum_l \rho_{j,l}\, v_l=\sum_l\left( \psi^*_j\, \psi_{l}\, +\delta_{j,l}\left(n_l-|\psi_l|^2\right)\right) v_l.
\end{equation}
Rearranging this equation and multiplying both sides by $\psi_j$ leads to
\begin{equation}
		\left(\lambda-n_j+|\psi_j|^2\right)\, \psi_j v_j=|\psi_j|^2\,\sum_l  \psi_{l} v_l.
\end{equation}
Finally, we divide both sides by $\lambda-n_j+|\psi_j|^2$ and sum over all lattice sites $j$ to obtain
\begin{equation}
\label{eigenvalue}
		1=\sum_j\frac{|\psi_j|^2}{\lambda-n_j+|\psi_j|^2}.
\end{equation}
This expression no longer contains the eigenvector $\mathbf{v}$ and therefore holds for all eigenvalues $\lambda$ of $\rho$. Defining $\alpha=\mbox{max}_j\{(n_j-|\psi_j|^2)\}$, one can directly infer that $\rho$ cannot have two different eigenvalues $\lambda$ and $\lambda'$ in the interval $(\alpha, N_{\mbox{\tiny tot}} ]$ both fulfilling Eq.~\ref{eigenvalue}. In this interval, all the addends in Eq.~\ref{eigenvalue} are positive and for $\lambda, \lambda' \in (\alpha, N_{\mbox{\tiny tot}} ]$ and $\lambda < \lambda'$ we find that
\begin{equation}
\label{eigenvalue2}
		1=\sum_j\frac{|\psi_j|^2}{\lambda-n_j+|\psi_j|^2} > \sum_j\frac{|\psi_j|^2}{\lambda'-n_j+|\psi_j|^2}=1
\end{equation}
leads to a contradiction.
 This reveals the interesting property, that from the intrinsic form of a bosonic Gutzwiller state, this cannot describe a fragmented BEC.

We now multiply Eq.~\ref{eigenvalue} by $\lambda$, which can then be seen as a fixed point, which is attractive in the interval $\lambda \in (\alpha, N_{\mbox{\tiny tot}} ]$. With this equation one can define the fixed point iteration 
\begin{equation}
		\lambda^{(i+1)}=\sum_j\frac{\lambda^{(i)} |\psi_j|^2}{\lambda^{(i)}-n_j+|\psi_j|^2}.
\end{equation}
This iteration, restricted to $\lambda^i \in (\alpha, N_{\mbox{\tiny tot}} ]$, always converges to the largest eigenvalue of the SPDM with arbitrary precision and directly scales with the number of lattice sites $L$. In most cases a relative error $\frac{\Delta \lambda}{\lambda}\leq 10^{-6}$ is already obtained after $3$-$15$ iterations.

\section{Explicit evaluation of the many-particle state at certain rational multiples of the harmonic oscillator time}

The aim of this section is to evaluate $|F(n)|^2$, physically corresponding to the quasi-momentum distribution of the condensate state at certain points in time after evolving in a harmonic confining potential, with 
\spl{
F(n)=\sum_{l=0}^{L-1}e^{2\pi i \frac{nl}{L}}\; e^{2\pi i \frac{l^2}{m}}.
}

Here $n,L\in \mathbb N$ are integers and $L$ is an integer multiple of $m$, i.e. $L/m\in \mathbb{N}$, to avoid broadening of the peaks in the resulting spectrum.

The summation integer $l$ is split into $l=p+qm$, where the newly defined integers can independently take on values in the ranges $p\in\{0,\ldots ,m-1\}$ and $q\in\{0,\ldots ,\frac{L}{m}-1\}$. Thus, the previous sum can be written as
\spl{
F(n)&=\sum_{p=0}^{m-1} \sum_{q=0}^{\frac{L}{m}-1}     e^{2\pi i \frac{n}{L}(p+qm)}\; e^{2\pi i \frac{(p+qm)^2}{m}}\\
&=  \sum_{q=0}^{\frac{L}{m}-1} e^{2\pi i \frac{m}{L}nq}       \sum_{p=0}^{m-1}     e^{2\pi i (\frac{np}{L}+\frac{p^2}{m}  )}
}
Using the identity 
\begin{equation}
	\sum_{n=0}^{N-1}e^{2\pi i \frac{k-k'}{N}n}=N\delta_{k,k'},
\end{equation}
for $N=\frac L m$, one obtains
\begin{equation}
F(n)=\frac L m \; \delta_{n \bmod \frac{L}{m},0} \;\sum_{p=0}^{m-1}     e^{2\pi i (\frac{np}{L}+\frac{p^2}{m}  )}.
\end{equation}

This restricts the resulting spectrum to a form with non-zero values only at $m$ periodically spaced positions $n=r \frac L m$ with $r\in \mathbb N$. To determine the value of the sum at these discrete values of $n$, it is convenient to define the function
\begin{equation}
	g(r)=\sum_{p=0}^{m-1} \; e^{2\pi i (\frac{rp}{m}+\frac{p^2}{m}  )}
\end{equation}
and it will now be shown that $|g(r)|^2$ is independent of $r\in \mathbb N$ for a fixed value of $m$.
\spl{
|g(r)|^2&=\sum_{p,p'=0}^{m-1} \; e^{2\pi i  (r\frac{p-p'}{m}+\frac{p^2-p'^2}{m}  )}\\
&=m+\sum_{p=1}^{m-1} \sum_{p'=0}^{p-1} \; \left[ e^{2\pi i (r\frac{p-p'}{m}+\frac{(p+p')(p-p')}{m}  ) }+ \mbox{h.c.}  \right].
}
We now perform a transformation of the summation indices and define $d=p-p'$ and $s=p+p'$. The summation over all terms can be reexpressed in the new variables as

\spl{
\sum_{p=1}^{m-1} \sum_{p'=0}^{p-1} \mapsto \sum_{d=1}^{m-1} \sum_{\stackrel{s=d}{\mbox{\tiny step }2}}^{2(m-1)-d}.
}

\begin{figure}[t!]
\includegraphics[width=0.4\linewidth]{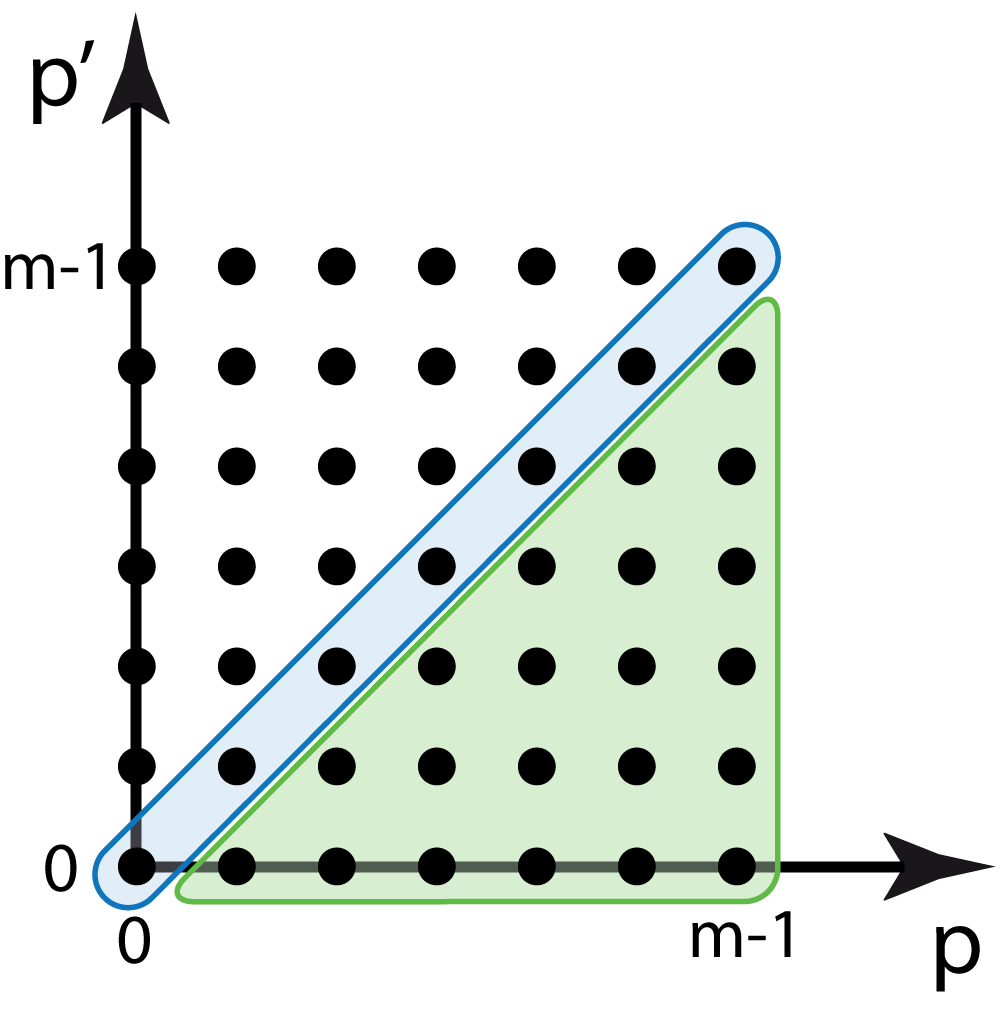}
\vspace{-0.3cm}
    \caption
    {\label{fig:resummation}
Illustration of the resummation procedure. The sum of the diagonal values amounts to $m$, while the values in the lower right triangle correspond to the complex conjugated values in the upper left triangle, mirrored on the diagonal. A summation over $s=p+p'$ at fixed $d=p-p'$ can be understood as sums of the $d$�th lower off-diagonal, where subsequent values of $s$ are spaced by $2$.
}
\end{figure}

Subsequently,
\spl{
\label{g_r_intermediate1}
|g(r)|^2&= m+  \sum_{d=1}^{m-1} \sum_{\stackrel{s=d}{\mbox{\tiny step }2}}^{2(m-1)-d}  \left[ e^{2\pi i (r\frac{d}{m}+\frac{sd}{m}  ) }+ \mbox{h.c.}  \right]\\
&=m+ \sum_{d=1}^{m-1} \left[ e^{2\pi i r\frac{d}{m} } \; h(d,m)  + \mbox{h.c.}  \right],
}

where we defined the function 
\spl{\label{h_fct_def}
h(d,m)&=\sum_{q=\frac d 2}^{m-1-\frac d 2} e^{2 \pi i \frac{2qd}{m}  }
}
with the summation variable $q=s/2$ now running over values with a step size of $1$. To evaluate this function for fixed $m$, the explicitly known expression for the geometric series can be used as long as $d\neq \frac m 2$. In the latter case, it can easily be evaluated and Eq.~\ref{h_fct_def} takes on a value $h(d=m/2,m)=(-1)^d \; d$, leading to the generally valid expression
\begin{widetext}
\spl{
h(d,m)&=(1-\delta_{d,m/2})\left(\frac{1-e^{4\pi i \frac d m (m-\frac d 2)}}{1-e^{4\pi i \frac d m }}     - \frac{1-e^{4\pi i \frac d m \frac d 2}}{1-e^{4\pi i \frac d m }} \right)+\delta_{d,m/2} (-1)^d \; d\\
&=(1-\delta_{d,m/2})\left(\frac{2 e^{-2\pi i \frac d m} \sin(2\pi  \frac {d^2}{m}) \sin(2\pi  \frac {d}{m}) }{ \cos(4\pi  \frac d m )-1}     \right)+\delta_{d,m/2} (-1)^d \; d.
}
\end{widetext}
Inserting this result into Eq.~\ref{g_r_intermediate1}, one obtains
\begin{widetext}
\spl{
\label{g_r_intermediate2}
|g(r)|^2&= m+  \delta_{0, m \bmod 2} \;  (-1)^{\frac m 2} \;  m  \, \cos(\pi r)+\sum_{d=1}^{m-1}(1-\delta_{d,m/2})\left(\frac{4 \cos(2\pi \frac d m (r-1)) \; \sin(2\pi  \frac {d^2}{m}) \; \sin(2\pi  \frac {d}{m}) }{ \cos(4\pi  \frac d m )-1}     \right)
}
\end{widetext}

The second term only contributes for even $m$, whereas the sum is identically equal to zero, as will now be shown. The term $(1-\delta_{d,m/2})$ ensures that only an even number of terms may contribute in the sum. This allows for a regrouping of terms and a corresponding resummation, where the terms $d=c$ and $d=m-c$ are considered as pairs, subsequently amounting to zero by using the periodicity and symmetry of the trigonometric functions:
\spl{
\label{g_r_intermediate2}
|g&(r)|^2= m+  \delta_{0, m \bmod 2} \;  (-1)^{\frac m 2 +r} \;  m  \\
&+\sum_{c=1}^{\lfloor \frac{m-1}{2} \rfloor}\left(\frac{4 \cos(2\pi \frac c m (r-1)) \; \sin(2\pi  \frac {c^2}{m}) \; \sin(2\pi  \frac {c}{m}) }{ \cos(4\pi  \frac c m )-1}\right.\\
  &+ \left. \frac{4 \cos(2\pi \frac {m-c} m (r-1)) \; \sin(2\pi  \frac {(m-c)^2}{m}) \; \sin(2\pi  \frac {m-c}{m}) }{ \cos(4\pi  \frac {m-c} m )-1}    \right)\\
  &= m (1+  \delta_{0, m \bmod 2} \;  (-1)^{\frac m 2 +r}).
}

We draw the connection to the quasi-momentum distribution from the condensate state in a harmonic trap at time $t$
\spl{
\label{nk_condensate}
n(\bk,t)&=\frac{1} L \sum_{l,l'} \phi_l(0) \phi_{l'}(0)^* \; e^{-i(ak-2tV_t l_0)(l-l') } \, e^{-it V_t(l^2-l'^2)}\\
&=\left| \frac{1}{\sqrt L} \sum_l   \phi_{l}(0) \; e^{-i(ak-2tV_t l_0)l } \, e^{-it V_t l^2} \right|^2,
}
where the initial condensate state is $ \phi_{l}(0)$. Assuming that the condensate is initially in a homogeneous state, i.e. $\phi_l(0)=\phi(0)=e^{i\varphi_0}L^{-\frac 1 2}$ with an irrelevant overall phase $\varphi_0$, the comparison of Eq.~\ref{nk_condensate} with $F(n)$ shows that at the specific times $t=\frac{2\pi}{m V_t}$ by equating 
\spl{
ak-\frac{4\pi}{m} l_0=2\pi \frac n L.
}
and using $r=\frac{mn}{L}$, one can relate

\spl{
\label{nk_of_t}n(\bk,t)&=\frac{|\phi(0)|^2}{L} \left|F\left(n=\frac{akl}{2\pi}-\frac{2l l_0}{m}\right)\right|^2\\
&=\frac{|\phi(0)|^2}{L} \frac{L^2}{m^2} \, \delta_{n \bmod \frac{L}{m},0} \left|g\left(\frac{akm}{2\pi}-{2 l_0}\right)\right|^2.
}
Thus, at the specific time $t=\frac{2\pi}{m V_t}$
\spl{
\label{nk_of_t_res}
n(\bk,t)&=\frac{1}{m} \delta_{\left(\frac{akm}{2\pi} - {2  l_0}\right) \bmod 1 \,,\, 0      } \\ 
& \times \left( 1+\delta_{0,m \bmod 2} (-1)^{\frac m 2 + \frac{akm}{2 \pi} - 2l_0  } \right).
}

If the condensate is initially not in a perfectly homogeneous state, but varies slowly in density, the peaks are broadened, but the underlying structure still remains.

For odd integer values of $m$, the term in brackets always takes on the value one and the momentum profile consists of an equidistant series of peaks at $k=\frac{2\pi}{am}(r+2l_0)$ and $r\in \mathbb Z$. As required from the unitarity of the Fourier transform, the quasi-momentum profile is of course normalized $\sum_k n(\bk,t=\frac{2\pi}{m V_t})=1$ at these certain times and each peak carries exactly the same weight $1/m$.

For even integer values of $m$, Eq.~\ref{nk_of_t_res} can be written as
\spl{
n(\bk,t)=\sqrt{\frac{2}{m}} \delta_{\left( \frac m 4 \left[1+\frac{ak}{\pi}\right] -  l_0 \right) \bmod 1 \, , \, 0      }
}
and the quasi-momentum profile consists of only $m/2$ equally strong and equidistantly spaced peaks at $k=\frac \pi a \left[\frac 4 m (s+l_0)-1 \right]$ with $s \in \mathbb Z$.

This analytical result agrees perfectly with the results of our numerical simulations. To the best of our knowledge, this analytical expression, which is also of direct relevance for the Talbot effect in quantum optics, has not been obtained so far.

\end{appendix}


\begin{thebibliography}{10}
\bibitem{N_central_peak}
We define the particle number in the projected 2D central peak as $N_p(t)=\sum_{k_z}\sum_{\|(k_x,k_y)\|_1 \leq 0.1 \frac{\pi}{a} } n(\bk,t)$.
\bibitem{thermal_fluct}
It is assumed that the time scale of the thermal expansion is large with respect to $1/V_t$.
\end{thebibliography}
\end{document}